# Modeling of Teager Energy Operated Perceptual Wavelet Packet Coefficients with an Erlang-2 PDF for Real Time Enhancement of Noisy Speech

Md Tauhidul Islam[a], Celia Shahnaz[b,*], Wei-Ping Zhu[c], M. Omair Ahmad[c]

[a]*Department of Electrical and Computer Engineering, Texas A&M University, College Station, Texas, USA-77840*
[b]*Department of Electrical and Electronic Engineering, Bangladesh University of Engineering and Technology, Dhaka-1000, Bangladesh*
[c]*Department of Electrical and Computer Engineering, Concordia University, Montreal, Quebec H3G 1M8, Canada*

**Abstract**

In this paper, for real time enhancement of noisy speech, a method of threshold determination based on modeling of Teager energy (TE) operated perceptual wavelet packet (PWP) coefficients of the noisy speech and noise by an Erlang-2 PDF is presented. The proposed method is computationally much faster than the existing wavelet packet based thresholding methods. A custom thresholding function based on a combination of $\mu$-law and semisoft thresholding functions is designed and exploited to apply the statistically derived threshold upon the PWP coefficients. The proposed custom thresholding function works as a $\mu$-law or a semisoft thresholding function or their combination based on the probability of speech presence and absence in a subband of the PWP transformed noisy speech. By using the speech files available in NOIZEUS database, a number of simulations are performed to evaluate the performance of the proposed method for speech signals in the presence of Gaussian white and street noises. The proposed method outperforms some of the state-of-the-art speech enhancement methods both at high and low levels of SNRs in terms of standard objective measures and subjective evaluations including formal listening tests.

**1. Introduction**

Determination of a signal that is corrupted by additive or multiplicative noise has been of interest because of its importance in both theoretical and practical fields. The main interest is to recover the signal from the noise-mixed data received from microphone, ECG machine, radar, mobile phone or any other sound devices. The use of such operation has application in broad area of speech communication applications, such as mobile telephony, speech coding and recognition, and hearing aid devices [9, 25, 28].

Several methods have been proposed in the last few decades to solve the problem of noise reduction and speech enhancement. Among these methods, time domain methods such as subspace approach [11, 15], frequency domain methods like discrete cosine transform based methods [6], spectral subtraction [5, 17, 18, 39], minimum mean square error (MMSE) estimator [11, 25], Weiner filtering [2, 4], compressive sensing based methods [13, 37, 38] and time-frequency domain methods such as wavelet or wavelet packet based thresholding methods [3, 10, 19, 23, 35] are the prominent ones. These methods have their own advantages and disadvantages. Subspace based approaches are computationally intensive and are not suitable for real time speech enhancement. Spectral subtraction based methods are very fast but have an intrinsic noise namely musical noise. MMSE and Wiener filtering based methods have moderate computational load but offer no mechanism to control the trade-off between speech distortion and noise reduction. Compressive sensing based methods applied in frequency domain are recent methods which have shown very promising results in reduction of noise while preserving the speech [13, 37, 38].

*Corresponding author
 *Email address:* celia.shahnaz@gmail.com (Celia Shahnaz)

The wavelet and wavelet packet based methods are superior choices from the perspective of controlling the speech distortion and noise reduction in comparison to time and frequency domain methods, but are computationally intensive. Universal thresholding method (UTM) [10], SURE [27], WPF [3] and BayesShrink [7] are prominent among the methods which use thresholding as process of removing noise. A common threshold is calculated in UTM and applied to the wavelet coefficients of noisy speech signal globally. Stein's uncertainty is applied in SURE method and Bayes principle is applied in BayesShrink method to determine the threshold. WPF is a modified version of universal thresholding method which has ability to detect the speech and silent frames. UTM is computationally much faster than other wavelet or wavelet packet based thresholding methods.

In the literature, employment of TE operator on the noisy speech or wavelet coefficients of the noisy speech is found to improve the discriminability of speech coefficients with noise coefficients [3, 21]. The TE operator is a powerful nonlinear operator proposed by Kaiser [24], capable of extracting the energy of a signal based on the mechanical and physical considerations. As TE operator computes instantaneous energy relative to its immediate neighbors, i.e., TE operator gives higher temporal resolution [29]. It has been successfully used in many speech processing applications [22, 30].

In this paper, instead of direct employment of the TE operator on the noisy speech, we apply the TE operator on the PWP coefficients of the noisy speech. The main contribution of this paper lies in the fitting of an Erlang-2 PDF suitable for modeling the TE operated PWP coefficients of noisy speech and then using it to determine analytically an appropriate subband-adaptive threshold. Erlang-2 PDF is a single parameter PDF that depends only on the variance of the modeled random variable and thus does not require computation of any other parameter for proper fit, which makes the proposed method significantly fast and suitable for real-time enhancement of noisy speech. We show that using Erlang-2 PDF in place of student $t$ PDF as in [19] does not degrade the performance of the speech enhancement procedure rather it improves the speed significantly. Designing a custom thresholding function based on the idea of combining $\mu$-law and semisoft thresholding functions is another contribution of the paper. This newly designed custom thresholding function is found to be more efficient in speech enhancement than the thresholding function based on combination of modified hard and semisoft used in [19]. The thresholding function proposed in this paper changes its characteristics based on the probabilities of speech presence and absence in a subband. The calculated threshold is applied on the PWP coefficients of noisy speech in order to obtain an enhanced speech by employing the proposed custom thresholding function.

The paper is organized as follows. Section 2 presents the proposed method of deriving an subband adaptive threshold and designing an appropriate thresholding function for enhancing noisy speech. Section 3 describes the detail of simulation and results with performance comparison. Concluding remarks are presented in Section 4.

## 2. Proposed method

Fig.1 shows the block diagram of the proposed method. We see from this figure that the noisy speech is framed and windowed at first and then a PWP transform is applied on this windowed frame, from where the PWP coefficients are obtained. TE operation is performed on these PWP coefficients to determine a proper threshold for employing the thresholding function. After application of the proposed custom thresholding function on the PWP coefficients using the threshold calculated in previous step, we obtain the PWP coefficients of an enhanced speech frame. The enhanced speech frame is obtained by inverse PWP transform of these PWP coefficients. Finally, enhanced speech is reconstructed by using the overlap-add method on the enhanced speech frames.

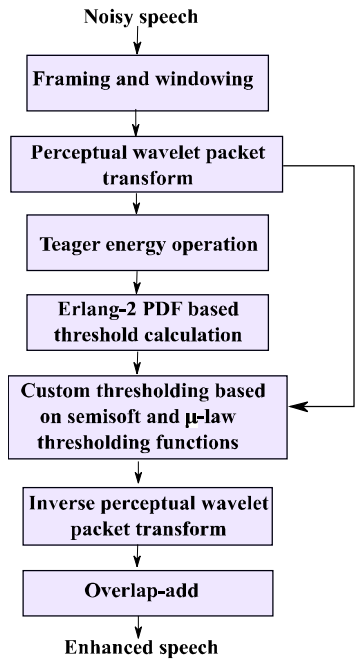

Figure 1: Block diagram of the proposed method

### 2.1. PWP transform and TE operation

Decomposition of the noisy speech according to the characteristics of human auditory system is the main motivation of PWP transform. A Mel warping function is used to determine the decomposition structure of the wavelet packets. Mel warping function is a nonlinear function of frequency. For this reason, the number of frequency bands in lower frequency is larger than the number of frequency bands in higher frequency in PWP transform. The motivation behind such a transform is that although human cochlea can differentiate the pitches precisely in lower frequencies, it cannot differentiate between small differences in higher frequencies. The PWP transform is discussed in detail in [34].

As discussed in [19], PWP transform does not provide enough frequency resolution to make differences between speech and noise PWP coefficients and TE operator can be a helpful tool to obtain the required frequency resolution. For this reason, we apply discrete time TE operator on the PWP coefficients in our proposed method.

Letting $W_{k,m}$ as the $m^{th}$ PWP coefficient in the $k^{th}$ subband, the TE operated coefficient $t_{k,m}$ corresponding to $W_{k,m}$ can be expressed as

$$t_{k,m} = T(W_{k,m}), \tag{1}$$

where discrete TE operator $T(W_{k,m})$ is defined as [24]

$$T(W_{k,m}) = W_{k,m}^2 - W_{k,m+1}W_{k,m-1}. \tag{2}$$

*2.2. Erlang-2 PDF as a model of TE operated PWP coefficients*

Determination of a proper threshold is the main concern in a thresholding based speech enhancement method. As used in [10], a unique threshold for all the subbands is not reasonable for thresholding in wavelet, wavelet packet or PWP domain [19, 31]. If the threshold can be determined based on the statistical properties of the coefficients in a subband, it is expected to be an accurate one for that subband. In our current context, if we can model the TE operated PWP coefficients of noisy speech and noise using a common PDF, we can use the theory of entropy between them to find out the exact threshold in a subband that defines the differentiating value between the speech and noise PWP coefficients. As discussed in [19] and [36], for time-varying nature of the speech signals, it becomes extremely difficult to realize the actual PDF of speech PWP coefficients or its $t_k$, where $t_k = t_{k,1}, \ldots, t_{k,M}$, $M$ is the total number of PWP coefficients in $k^{th}$ subband. In place of formulating a PDF of $t_k$, we can formulate the histogram of $t_k$ and can find a PDF that closely resembles the histogram. As mentioned in [19], Student $t$ PDF is a good choice for approximating the histogram. But as determination of the proper degree of freedom in the Student $t$ PDF is time-consuming, this method is not suitable for real time speech enhancement. For this reason, we looked for a PDF that depends only on the variance of $t_k$ of the noise or noisy speech it models and closely resembles the histogram. Erlang-2 PDF can be an option as it depends only on the variance of the model it fits. To test this, we superimpose the empirical histogram along with Gaussian, Student $t$ and Erlang-2 PDFs for TE operated PWP coefficients in a randomly chosen subband of a noisy speech frame in Figs. 2, 3 and 4 in presence of Gaussian white noise at SNRs of −15, 0 and 15 dB. From these figures, it is obvious that Erlang-2 PDF performs very similar to the Student $t$ PDF, but has a better fit with the empirical histogram than Gaussian PDF. We obtain similar analysis for empirical histogram, Gaussian, Student $t$ and Erlang-2 PDFs of TE operated noise PWP coefficients at the same subband of the same noisy speech frame in presence of Gaussian white noise at the same SNRs as used in Figs. 2, 3 and 4 which are shown in Figs. 5, 6 and 7. Such matching of the empirical histogram with Erlang-2 PDF in comparison with the Gaussian and Student $t$ PDFs can also be explained in terms of AIC index [1]. It can be noted from [1] that the more negative value of the AIC index indicates more matching between the data and PDF model. Assuming Gaussian, Student $t$ and Erlang-2 PDFs for TE operated noise PWP coefficients in 1000 randomly chosen subbands of a noisy speech, mean values of AIC indices with standard deviations obtained are shown in Fig. 8 at an SNR range of −15dB to 15dB in the presence of Gaussian white noise. From Fig. 8, it is clearly seen that the Erlang-2 PDF offers similar matching to Student $t$ PDF, but better matching with the empirical histogram compared to the Gaussian PDF not only at SNR of 15dB but also at an SNR as low as −15dB. The plot representing the mean values of AIC indices with standard deviations for the Gaussian, Student $t$ and Erlang-2 PDFs of TE operated noise PWP coefficients in the same 1000 subbands of the same noisy speech at SNR level ranging from −15dB to 15dB is illustrated in Fig. 9. This figure shows that mean AIC indices for $t_k$ of noise continues to exhibit similar values for Erlang-2 and Student $t$ PDFs , but smaller values for Erlang-2 PDF than the Gaussian PDF thus maintaining better matching with the empirical histogram for a wide range of SNR. Therefore, we propose to approximate the histograms of TE operated PWP coefficients of noisy speech and noise by Erlang-2 PDF and use this approximation to calculate the threshold adaptive to different PWP subbands.

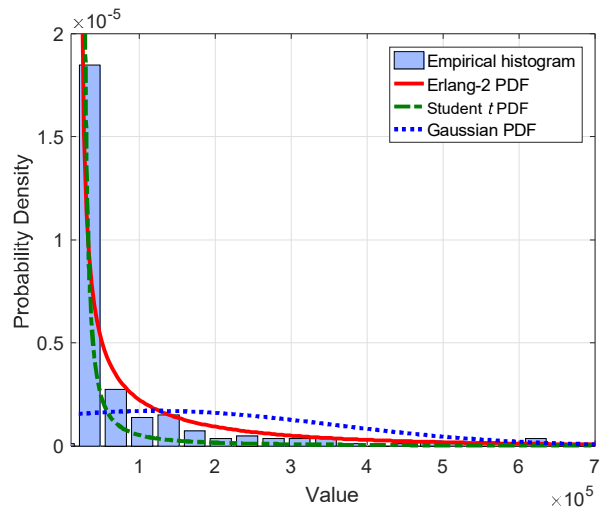

Figure 2: Empirical histogram, Erlang-2, Student $t$ and Gaussian PDFs of TE operated PWP coefficients of noisy speech at SNR of −15 dB

*2.3. Determination of the subband-adaptive threshold*

The entropy between TE operated noisy speech PWP coefficients and TE operated noise PWP coefficients is different in each subband based on the speech and noise power in that subband [19]. Therefore, measurement of entropy between them can be a useful way to choose a threshold value adaptive to each subband. Some popular similarity measures those can be considered as measurement of the entropy are variational distance, Bhattacharyya distance, harmonic mean, Kullback Leibler (K-L) divergence, and symmetric K-L divergence. As K-L divergence is computationally less demanding and a good choice for statistical entropy measure, K-L divergence and the symmetric K-L divergence are two good choices in our current context for measurement of entropy between noisy speech PWP coefficients and noise PWP coefficients. The K-L divergence is always non-negative and zero if and only if the approximate Erlang-2 PDF of $t_k$ of noisy speech and that of noise or the approximate Erlang-2 PDF of $t_k$ of the noisy speech and that of the clean speech are exactly the same. But The K-L divergence is not symmetric with the two PDFs. For this reason, in order to have a symmetric distance between PDFs of $t_k$ of PWP coefficients of noisy speech, clean speech and noise, we adopt the symmetric K-L divergence in this paper. The symmetric K-L divergence is defined as

$$SKL(p,q) = \frac{KL(p,q) + KL(q,p)}{2}, \qquad (3)$$

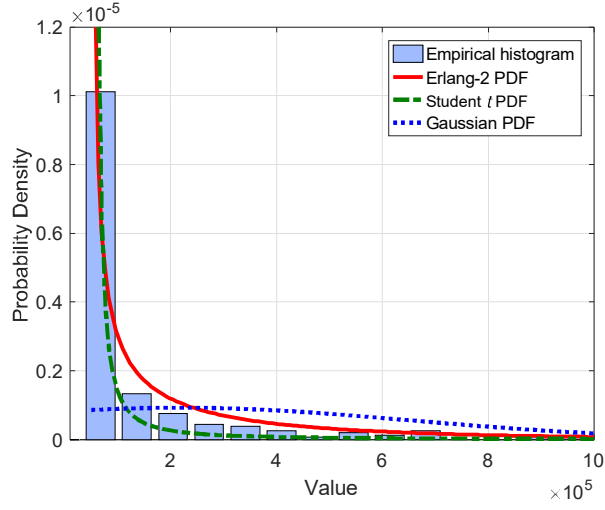

Figure 3: Empirical histogram, Erlang-2, Student *t* and Gaussian PDFs of TE operated PWP coefficients of noisy speech at SNR of 0 dB

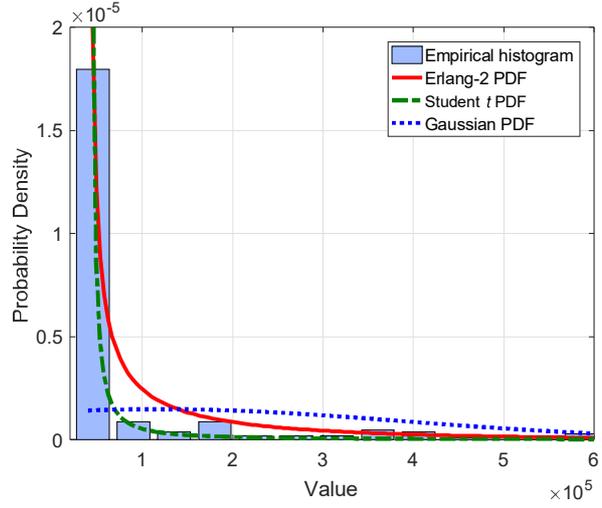

Figure 4: Empirical histogram, Erlang-2, Student *t* and Gaussian PDFs of TE operated PWP coefficients of noisy speech at SNR of 15 dB

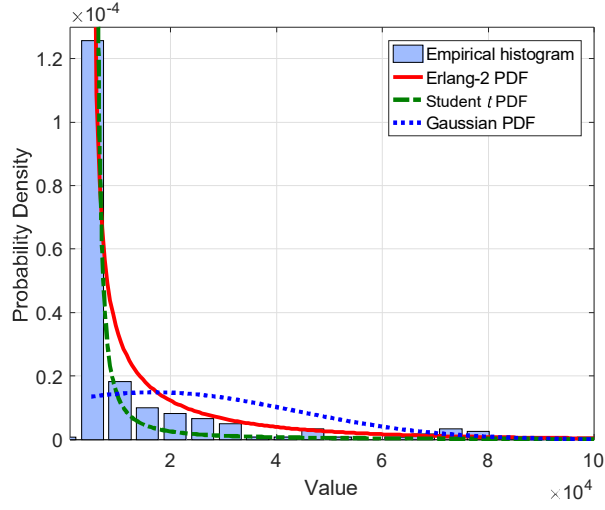

Figure 5: Empirical histogram, Erlang-2, Student $t$ and Gaussian PDFs of TE operated noise PWP coefficients at SNR of $-15$ dB

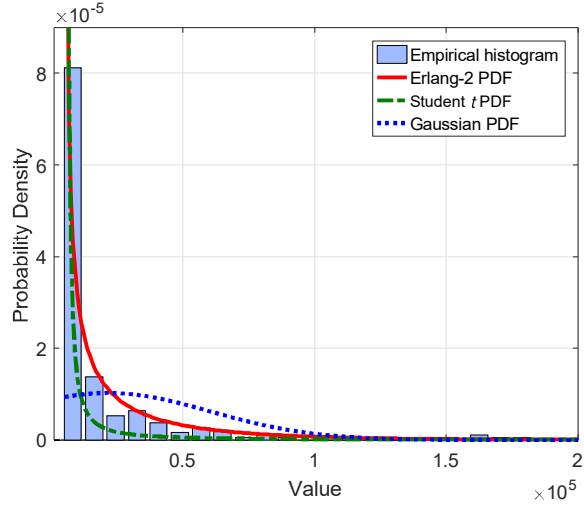

Figure 6: Empirical histogram, Erlang-2, Student $t$ and Gaussian PDFs of TE operated noise PWP coefficients at SNR of 0 dB

where $p$ and $q$ are two PDFs calculated from the corresponding histograms each having $N$ number of bins and $KL(p,q)$ is the K-L divergence between $p$ and $q$ given by

$$KL(p,q) = \sum_{i=1}^{N} p_i(t_k) \ln \frac{p_i(t_k)}{q_i(t_k)}. \qquad (4)$$

In (4), $p_i(t_k)$ is the probability of $t_k$ of noisy speech in $i^{th}$ bin given by

$$p_i(t_k) = \frac{n_i}{M}, \qquad (5)$$

where $n_i$ is number of coefficients in $i^{th}$ bin and $M$ total number of coefficients in $k^{th}$ subband. In this way, we can find the PDF for $t_k$ of noisy speech, $p_{t_k}$ and approximate the obtained PDF with Erlang-2 PDF. We will denote the approximated Erlang-2 PDF with $\hat{p}_{t_k}$. Similarly, the approximate Erlang-2 PDF of $t_k$ of the noise can be estimated following (5) and denoted by $\hat{q}_{t_k}$. Below a certain value of threshold $\lambda$, the symmetric K-L divergence between $\hat{p}_{t_k}$ and $\hat{q}_{t_k}$ is approximately zero, i.e.,

$$SKL(\hat{p}_{t_k}, \hat{q}_{t_k}) \approx 0. \qquad (6)$$

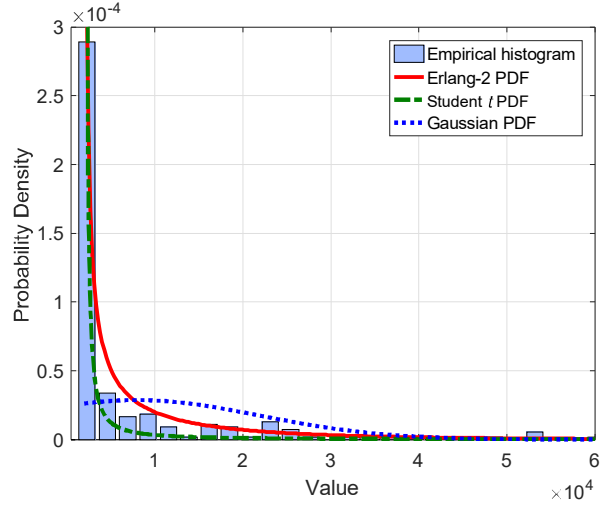

Figure 7: Empirical histogram, Erlang-2, Student *t* and Gaussian PDFs of TE operated noise PWP coefficients at SNR of 15 dB

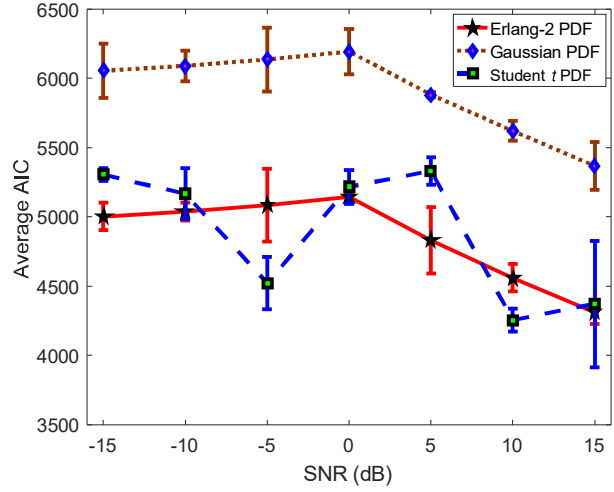

Figure 8: Mean values of AIC indices with standard deviations for TE operated PWP coefficients in 1000 subbands of noisy speech assuming Erlang-2, Student *t* and Gaussian PDFs

Erlang-2 PDF for $t_k$ of noisy speech can be written as [12]

$$\hat{p}_{t_k}(x) = \frac{2}{\sigma_s^2} x e^{-\sqrt{\frac{2}{\sigma_s^2}} x}, \qquad (7)$$

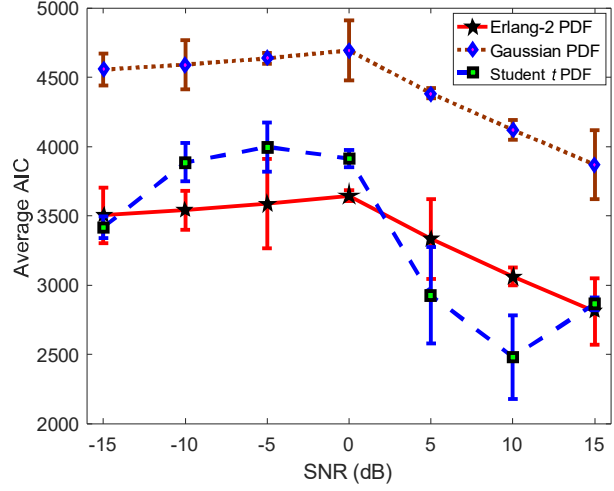

Figure 9: Mean values of AIC indices with standard deviations for TE operated noise PWP coefficients in 1000 subbands of noisy speech assuming Erlang-2, Student $t$ and Gaussian PDFs

where $\sigma_s^2$ represents the power of $t_k$ of noisy speech. Following (7), Erlang-2 PDF for $t_k$ of noise can be written as

$$\hat{q}_{t_k}(x) = \frac{2}{\sigma_n^2} x e^{-\sqrt{\frac{2}{\sigma_n^2}}x}. \tag{8}$$

By substituting (7) and (8) in (6), we obtain

$$\int_0^\lambda [\frac{2}{\sigma_s^2} x e^{-\sqrt{\frac{2}{\sigma_s^2}}x} - \frac{2}{\sigma_n^2} x e^{-\sqrt{\frac{2}{\sigma_n^2}}x}] I_1 dx = 0, \tag{9}$$

where $I_1$ is defined as,

$$I_1 = \ln(\frac{\sigma_n^2}{\sigma_s^2}) \times e^{-\sqrt{\frac{2}{\sigma_s^2}}x + \sqrt{\frac{2}{\sigma_n^2}}x}.$$

By solving (9), value of $\lambda(k)$ can be derived as

$$\lambda(k) = \frac{\sqrt{\sigma_n^2(k)\gamma_k} \ln \gamma_k}{\sqrt{\gamma_k} - 1}, \tag{10}$$

where $\gamma_k$ is the a posteriori signal to noise ratio (SNR) at subband $k$ defined as

$$\gamma_k = \frac{\sigma_s^2(k)}{\sigma_n^2(k)}. \tag{11}$$

The calculated threshold $\lambda(k)$ in (10) derived assuming Erlang-2 PDF is compared with that obtained assuming Gaussian PDF given by [31]

$$\lambda(k) = \sqrt{\sigma_n^2(k)} \sqrt{2(\gamma_k + \gamma_k^2)} \times \ln(\sqrt{1 + \frac{1}{\gamma_k}}) \tag{12}$$

and assuming Student $t$ PDF given by [19]

$$\lambda(k) = \sqrt{\frac{\sigma_n^2(k)(1 + \gamma(k))}{(\sqrt{1 + \gamma(k)} + 2 + \gamma(k))\sqrt{\chi_2}}}, \tag{13}$$

in Fig. 11 for $\chi_2 = 0.5$ and unit signal power. This figure shows that the pattern of the threshold value is similar for all three PDFs with respect to SNR level, i.e., the threshold is high for low SNRs and low for high SNRs. In terms of value, Erlang-2 PDF produces higher threshold value than Student $t$ PDF but lower value than Gaussian PDF at low SNRs as −15 dB. For SNR of more than −10 dB, Erlang-2 PDF shows threshold values those are significantly higher than both of the Student $t$ and Gaussian PDFs. Therefore, the threshold derived from the Erlang-2 PDF offers more chance of removing noise coefficients at high SNRs and preserving speech coefficients at low SNRs.

Since higher variance of $t_k$ of PWP coefficients of speech, noise and noisy speech indicates a higher variance of PWP coefficients of speech, noise and noisy speech, respectively, the proposed subband-adaptive threshold, derived in (10) is high for higher noise power and low for lower noise power in each subband thus is adaptive to noise power at different subbands. A voice activity detector (VAD) is not necessary in the proposed method because the threshold can automatically adapt to the silent and speech frames. In all the subbands of a silent frame, since noise power is significantly higher than the power of speech, the proposed threshold results in a higher value calculated from (10). Such a large value of the threshold imposes most of the coefficients in the subbands of that frame to be thresholded to zero thus removes noise coefficients completely. In the proposed method, we use the improved minima controlled recursive averaging (IMCRA) method [8] for estimating the noise.

*2.4. Thresholding function based on speech presence probability*

A detail discussion of the advantages of the semisoft and $\mu$-law thresholding functions over soft and hard thresholding functions is given in [33] with a proposal of new thresholding function based on their combination. But in [33], application of semisoft thresholding function is limited only in the subbands of silence frames and $\mu$-law thresholding function in speech subbands, which could not aggregate the true advantages of combination of these two methods. Here, we propose a custom thresholding function derived from $\mu$-law and semisoft thresholding functions, that truly encodes the advantages of their combination and behaves as a $\mu$-law or semisoft based on the probabilities of speech presence and absence. Representing $\lambda(k)$ derived from (10) using the variance of estimated noise coefficients and variance of PWP coefficients as $\lambda_1(k)$ and letting $\lambda_2(k) = 2\lambda_1(k)$, the proposed thresholding function can be expressed as

$$(Y_{k,m})_{PCT} = \begin{cases} \alpha(k,m)\frac{sgn(Y_{k,m}) \cdot |Y_{k,m}|}{\mu}[(1+\mu)^{\frac{|Y_{k,m}|}{\lambda_1(k)}} - 1], & \text{if } |(Y_{k,m})| \le \lambda_1(k), \\ Y_{k,m}, & \text{if } |(Y_{k,m})| \ge \lambda_2(k), \\ (1-\alpha(k,m))\Omega_1 + \alpha(k,m)\Omega_2, & \text{otherwise}, \end{cases} \qquad (14)$$

where

$$\Omega_1 = sgn(Y_{k,m}) \times \lambda_2(k) \frac{|(Y_{k,m})| - \lambda_1(k)}{\lambda_2(k) - \lambda_1(k)}, \qquad (15)$$

$$\Omega_2 = Y_{k,m}. \qquad (16)$$

In (14), $\mu$ is a constant that can be determined empirically for best performance of the proposed method and *sgn* represents the signum function. $(Y_{k,m})_{PCT}$ stands for the PWP coefficients thresholded by the proposed custom thresholding function expressed and shape parameter of the proposed thresholding function is represented by $\alpha(k,m)$ which is taken as a constant in the proposed method.

We compare the proposed thresholding function for $\alpha(k,m) = 0.5$ with the conventional $\mu$-law and semisoft thresholding functions in Fig. 10. Depending on the value of shape parameter $\alpha(k,m)$, it can be verified from (14) that the proposed thresholding function gets the following forms,

$$\lim_{\alpha(k,m)\to 0} (Y_{k,m})_{PCT} = (Y_{k,m})_{SS},$$

$$\lim_{\alpha(k,m)\to 1} (Y_{k,m})_{PCT} = (Y_{k,m})_{ML},$$

where $(Y_{k,m})_{MH}$ stands for the PWP coefficients thresholded by the modified hard thresholding function and $(Y_{k,m})_{SS}$ represents the PWP coefficients thresholded by the semisoft thresholding function.

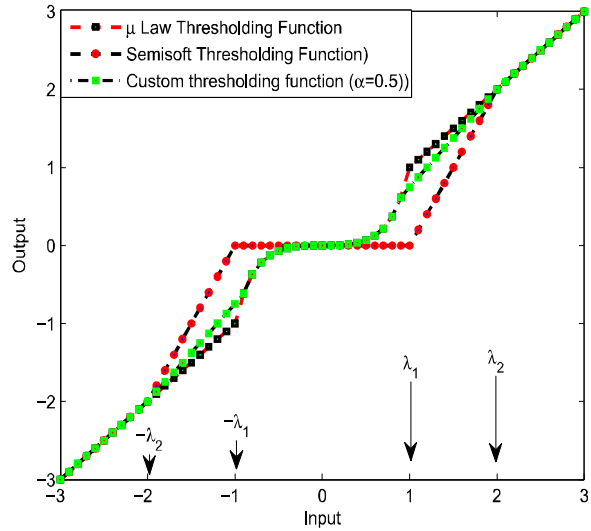

Figure 10: Input Output Relation for semisoft, $\mu$ law and proposed custom thresholding function

### 2.4.1. Determination of shape parameter

The shape parameter $\alpha(k,m)$ determines whether the thresholding function will behave like semisoft or $\mu$-law thresholding function. Following [19], we propose to determine $\alpha(k,m)$ as

$$\alpha(k,m) = \frac{1 + r(k,m)}{2(1 + q(k,m))}, \tag{17}$$

where $r(k,m)$ and $q(k,m)$ are the probabilities of the presence and absence of speech, respectively, for the $m^{th}$ coefficient in the $k^{th}$ subband and they are related to each other through

$$r(k,m) = 1 - q(k,m). \tag{18}$$

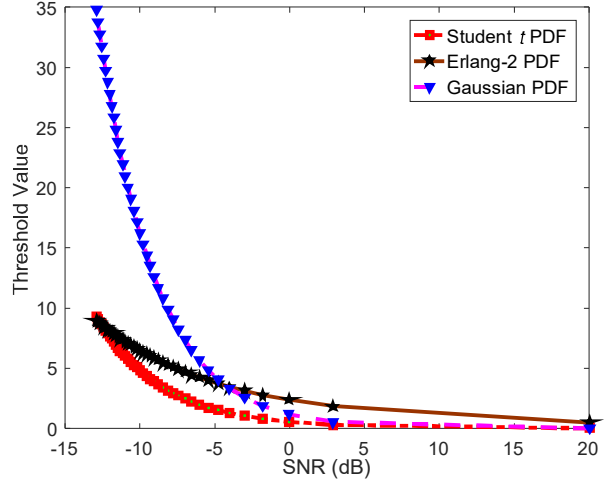

Figure 11: Comparison of threshold values with respect to SNR for Erlang-2, Student $t$ and Gaussian PDFs

When $r(k,m)$ is large, it means that the PWP coefficient is most likely a speech coefficient. On the other hand, a large value of $q(k,m)$ indicates that the PWP coefficient is most likely a noise coefficient. The probability of speech absence $q(k,m)$ in (18) is calculated by

$$q(k,m) = 1 - r_{local}(k,m) r_{global}(k,m) r_{subband}(k,m), \qquad (19)$$

where $r_{local}(k,m)$ and $r_{global}(k,m)$ are the probabilities of presence of speech in local and global windows in a subband. If we use $\tau$ to represent "local" or "global" window, $r_\tau(k,m)$ can be given by

$$r_\tau(k,m) = \begin{cases} 0, & \text{if } \xi_\tau(k,m) \leq \xi_{min} \\ 1, & \xi_\tau(k,m) \geq \xi_{max}, \\ \frac{\log(\xi_\tau(k,m)/\xi_{min})}{\log(\xi_{max}/\xi_{min})}, & \text{otherwise,} \end{cases} \qquad (20)$$

where $\xi_\tau(k,m)$ represents either "local" or "global" mean values of the *apriori* SNR. $\xi_\tau(k,m)$ is calculated by

$$\xi_\tau(k,m) = \sum_{i=-w_\tau}^{i=w_\tau} h_\tau(i) \xi(k-i,m), \qquad (21)$$

where $h_\tau$ is a normalized rectangular window of size $2w_\tau + 1$ and $\xi(k,m)$ denotes a recursive average of the *apriori* SNR [8]. $\xi(k,m)$ has an expression of

$$\xi(k,m) = \kappa \xi(k,m-1) + (1-\kappa) \widehat{\eta}(k,m-1), \qquad (22)$$

where $\kappa$ denotes an averaging constant and $\xi_{min}$ and $\xi_{max}$ are two empirically determined constants denoting minimum and maximum values of $\xi(k,m)$. $r_{subband}(k)$ in (19) can be calculated as

$$r_{subband}(k) = \begin{cases} 0, & \text{if } \xi_{subband}(k) < \xi_{min} \\ 1, & \text{if } \xi_{subband}(k) > \xi_{subband}(k-1) \text{ and} \\ & \xi_{subband}(k) > \xi_{min}, \\ \mu(k), & \text{otherwise,} \end{cases} \quad (23)$$

where expression of $\mu(k)$ can be written as

$$\mu(k) = \begin{cases} 0, & \text{if } \xi_{subband}(k) \leq \xi_{peak}(k)\xi_{min}, \\ 1, & \text{if } \xi_{subband}(k) \geq \xi_{peak}(k)\xi_{max}, \\ \frac{\log(\xi_{subband}(k)/\xi_{peak}(k)/\xi_{min})}{\log(\xi_{max}/\xi_{min})}, & \text{otherwise.} \end{cases} \quad (24)$$

In (23) and (24), $\xi_{subband}(k)$ is calculated as

$$\xi_{subband}(k) = \frac{1}{M} \sum_{1 \ll m \ll M} \xi(k,m) \quad (25)$$

and $\xi_{peak}$ in (23) is a confined peak value of $\xi_{subband}(k)$. Using the values of $r(k,m)$ and $q(k,m)$ determined from (18) and (19), we can calculate $\alpha(k,m)$ from (17).

*2.5. Inverse PWP transform*

For a noisy speech frame, we obtain thresholded PWP coefficients by calculating the proposed subband-adaptive threshold in (10) and using that threshold in the proposed custom thresholding function in (14). An enhanced speech frame $\widehat{r}[n]$ is synthesized by performing inverse PWP transform as $\widehat{r}[n] = PWP^{-1}(Y_{k,m})_{PCT}$. A standard overlap-add method is used to reconstruct the enhanced speech signal [28].

## 3. Results

In this section, we perform a number of simulations to evaluate the performance of the proposed method.

*3.1. Simulation conditions*

Speech files from the NOIZEUS database are used for the experiments used in this simulation, where the speech data are sampled at 8 KHz [14]. Noise sequences are added to the clean speeches at different SNR levels ranging from 15 dB to -15 dB to create the noisy speeches. Following [26], two different types of noise, namely white and street are adopted in this simulation [14].

Hamming windowing operation is performed for obtaining the overlapping analysis frames, where the size of each of the frame is taken as 640 samples with 50% overlap between two successive frames. A 6-level PWP decomposition with Daubechies 10 (db10) wavelet function is applied on the noisy speech frames resulting in 24 PWP subbands [34], [32] for each input frame. The values of the constants used to determine the shape parameter in the proposed thresholding function are given in Table 1. The value of $\mu$ is taken as 0.9.

Table 1: Constants used to determine the shape parameter

| Constants | Value |
|---|---|
| $\beta$ | 0.7 |
| $\xi_{min}$ | -10 dB |
| $\xi_{max}$ | -5 dB |
| $\xi_{peak}$ | 10 dB |
| $w_{local}$ | 1 |
| $w_{global}$ | 15 |

*3.2. Comparison metrics*

For evaluating the performance of the proposed method with some state-of-the art methods, namely, universal thresholding method (UTM) [10], soft mask estimator with posteriori SNR uncertainty (SMPO) [26] and Student *t* modeling based thresholding (STMT) [19], we take segmental SNR (SNRSeg) improvement, perceptual evaluation of speech quality (PESQ) and weighted spectral slope (WSS) as the comparison metrics [16]. The spectrograms are presented and formal listening tests are performed for subjective evaluation of the proposed method with the competing methods.

*3.3. Objective evaluation*

*3.3.1. Results for speech signals corrupted by Gaussian white noise*

SNRSeg improvement, PESQ and WSS for speech signals corrupted with Gaussian white noise for UTM, SMPO, STMT and proposed method are shown in Fig.12, Table 2 and Fig.13. In Fig. 12, we compare the performance of the proposed method with that of the other methods at different levels of SNR for Gaussian white noise corrupted speech in terms of SNRSeg improvement. We can realize from this figure that the SNRSeg improvement increases as SNR decreases. At a low SNR of $-15dB$, the proposed method yields the highest SNRSeg improvement of 3.98 dB which is comparable to STMT and much higher than other competing methods. Such a high value of SNRSeg improvement at a low level of SNR clearly attest the efficacy of the proposed method in producing enhanced speech with better quality for speech severely corrupted by Gaussian white noise. For the high SNR cases also, we see that the proposed method shows similar performance as STMT but better performance than UTM and SMPO.

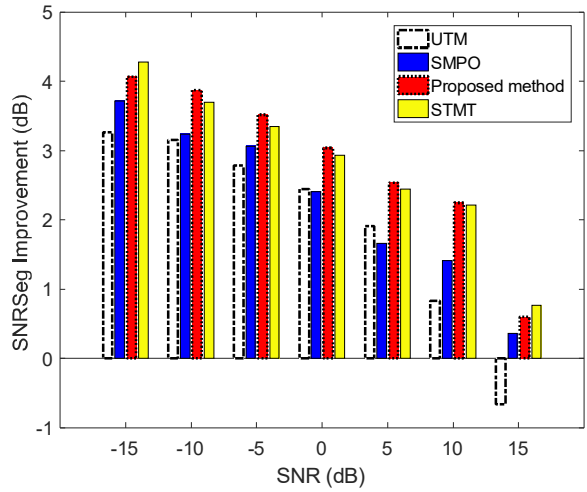

Figure 12: SNRSeg Improvement for different methods in Gaussian white noise

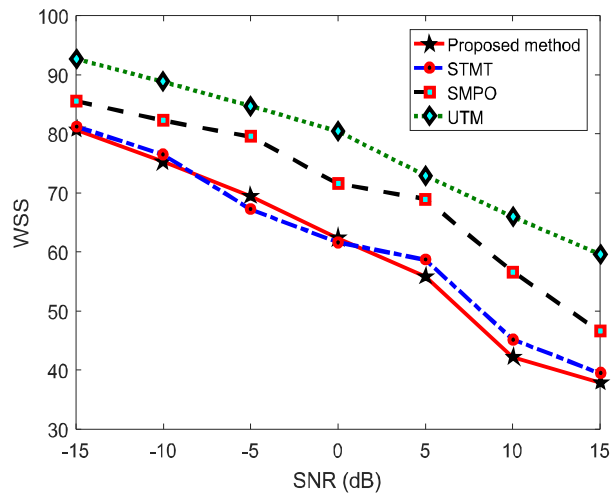

Figure 13: WSS for different methods in Gaussian white noise

PESQ scores for the proposed method with other competing methods is shown in Table 2. The proposed method consistently yields similar PESQ to STMT and higher PESQ at all the SNR levels in comparison to UTM and SMPO. Since PESQ score is the indicator of the speech quality, this table clearly indicates the capability of the proposed method in producing speech with better quality.

In Fig. 13, the WSS values as a function of SNR for the proposed method and those for the other competing methods are shown. From this figure, we can realize that the WSS values resulting from the competing methods except STMT are relatively larger for every levels of SNR levels in comparison to the proposed method. As lower WSS values indicate a better speech quality, the proposed method indeed is superior to UTM and SMPO.

Table 2: PESQ for different methods in presence of Gaussian white noise

| SNR(dB) | UTM | SMPO | STMT | Proposed method |
|---|---|---|---|---|
| -15 | 1.22 | 1.39 | 1.50 | 1.45 |
| -10 | 1.31 | 1.47 | 1.59 | 1.62 |
| -5 | 1.49 | 1.52 | 1.85 | 1.92 |
| 0 | 1.52 | 1.89 | 1.93 | 2.03 |
| 5 | 1.89 | 2.56 | 2.62 | 2.57 |
| 10 | 2.19 | 2.88 | 2.81 | 2.89 |
| 15 | 2.44 | 3.01 | 3.06 | 3.05 |

*3.3.2. Results for speech signals corrupted by street noise*

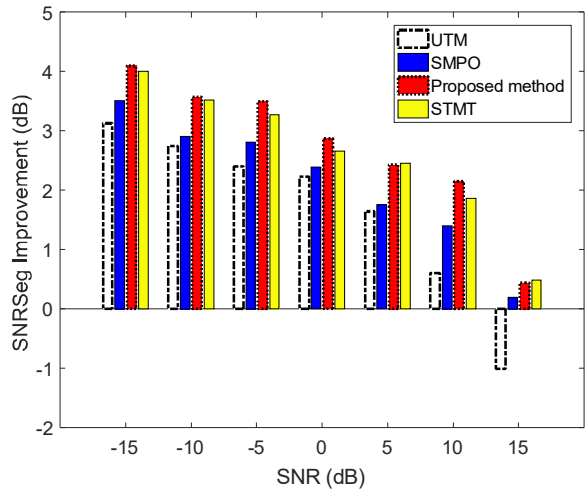

Figure 14: SNRSeg Improvement for different methods in street noise

SNRSeg improvement, PESQ and WSS for speech signals corrupted with street noise for universal thresholding, SMPO, STMT and proposed method are shown in Fig. 14, Table 3 and Fig. 15, respectively.

In Fig.14, it can be seen that at a low level of SNR of $-15dB$ of street noise corrupted speech, the proposed method provides a SNRSeg improvement that is close to STMT and significantly higher than those of UTM and SMPO, which is true for high SNRs also.

For speech signal corrupted with street noise, the values of PESQ obtained using the proposed method are compared with that of the other methods in Table 3. From this table, we can say that over the whole SNR range considered, the proposed method provides higher PESQ scores in the presence of street noise in comparison to other methods. For this case, even the proposed method outperforms STMT at all SNR levels.

The performance of the proposed method is compared with that of the other methods in terms of WSS in Fig.15

Table 3: PESQ for different methods in presence of street noise

| SNR(dB) | UTM | SMPO | STMT | Proposed method |
|---|---|---|---|---|
| -15 | 1.16 | 1.15 | 1.30 | 1.39 |
| -10 | 1.23 | 1.37 | 1.42 | 1.51 |
| -5 | 1.32 | 1.51 | 1.80 | 1.81 |
| 0 | 1.43 | 1.69 | 1.83 | 1.97 |
| 5 | 1.69 | 2.07 | 2.54 | 2.58 |
| 10 | 1.93 | 2.38 | 2.65 | 2.71 |
| 15 | 2.14 | 2.60 | 2.89 | 2.96 |

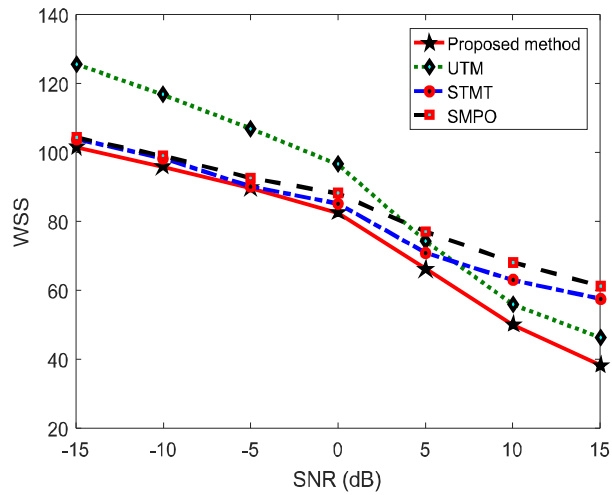

Figure 15: WSS for different methods in street noise

at different levels of SNRs in presence of street noise. It is clearly seen from this figure that WSS increases as SNR decreases. At a low SNR of $-15dB$, the proposed method yields a WSS that is lower than that of all other methods, which remains lower over the higher SNRs also. This case is another example where the proposed method outperforms STMT at all SNR levels.

*3.4. Results for speech signals corrupted by car and multi-talker babble noises*

The proposed method is also evaluated in presence of car and multi-talker babble noises through computing the SNRSeg improvements at different SNR levels of input noisy speech signals. The SNRSeg improvement for the proposed method with comparison methods for car noise is shown in Fig. 16 and for multi-talker babble noise in Fig. 17. In these two figures, we see the similar trend of efficacy in speech enhancement by different methods for car and multi-talker babble noises as street and Gaussian white noises. The proposed method is found to perform better or comparable in comparison to STMT for most of the SNR levels. The proposed method performs much better than UTM and SMPO for all levels of SNR.

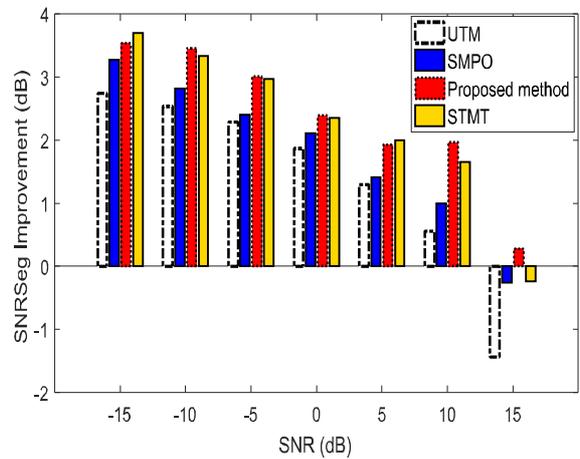

Figure 16: SNRSeg Improvement for different methods in car noise

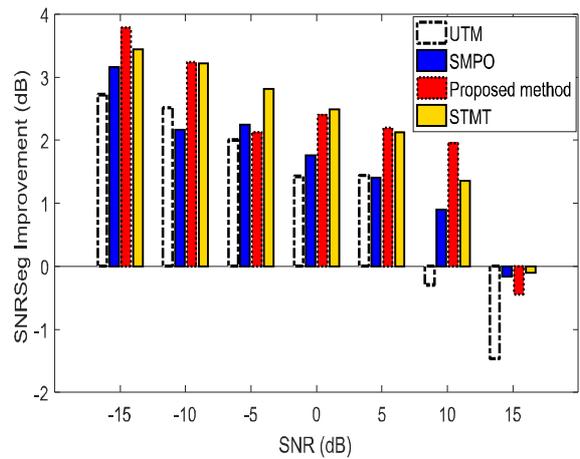

Figure 17: SNRSeg Improvement for different methods in babble noise

### 3.5. Subjective evaluation

For evaluation of the proposed method with other competing methods from the perspective of subjective evaluation, we use two tools. The first one is the plot of the spectrograms of the output for all the methods and compare their performance in terms of preservation of harmonics and presence of noise.

we plot the spectrograms of the clean speech, the noisy speech, and the enhanced speech signals obtained by using the proposed method and all other methods are presented in Fig.18 for white noise corrupted speech at an SNR of 10 dB. It is obvious from the figures that the proposed method preserves the harmonics similar to STMT but significantly better than UTM and SMPO. The noise is also reduced at every time point for the proposed method which attest our claim of better performance in terms of higher SNRSeg improvement, higher PESQ and lower WSS values in objective evaluation. Another collection of spectrograms for the proposed method with other methods for speech signals corrupted with street noise is shown in Fig. 19. This figure also attests that our proposed method has better performance in terms of harmonics' preservation and noise removal. The second tool we used for subjective evaluation of the proposed method and the competing methods is the formal listening tests. We add white and street noises to all the thirty speech sentences of NOIZEUS database at −15 to 15 SNR levels and process them with all the competing methods. We allow ten listeners to listen to these enhanced speeches from these methods and evaluate them subjectively. Following [19] and [14], We use SIG, BAK and OVL scales on a range of 1 to 5. The detail of these scales and procedure of this listening test is discussed in [19]. More details on this testing methodology of listening test can be obtained from [20].

Table 4: Mean Scores of SIG scale for different methods in presence of Gaussian white noise at 5 db

| Listener | UTM | SMPO | STMT | Proposed Method |
|---|---|---|---|---|
| 1 | 3.6 | 4.0 | 4.1 | 4.0 |
| 2 | 3.3 | 3.9 | 3.7 | 3.7 |
| 3 | 3.9 | 4.0 | 4.3 | 4.2 |
| 4 | 3.4 | 4.2 | 4.2 | 4.5 |
| 5 | 3.2 | 3.8 | 4.1 | 4.0 |
| 6 | 2.9 | 4.6 | 4.9 | 4.9 |
| 7 | 3.8 | 3.8 | 4.3 | 4.2 |
| 8 | 3.5 | 3.7 | 4.2 | 4.2 |
| 9 | 3.5 | 4.5 | 4.7 | 4.8 |
| 10 | 3.7 | 4.8 | 4.9 | 4.8 |

We show the mean scores of SIG, BAK, and OVRL scales for all the methods for speech signals corrupted with Gaussian white noise in Tables 4, 5, and 6 and for speech signals corrupted with street noise is shown in Tables 7, 8, and 9. The higher values for the proposed method in comparison to other methods clearly attest that the proposed method is better than them in terms of lower signal distortion (higher SIG scores), efficient noise removal (higher BAK scores) and overall sound quality (higher OVL scores) for all SNR levels.

*3.6. Comparison of computational time*

The proposed method with all the competing methods are run in a personal computer with Intel core-i7 processor and 16-GB RAM for a speech file of 3s. Table 10 shows the required time for each method to process the speech file for different window lengths. From this table it is clearly seen that the proposed method is computationally 10000 times faster than STMT and comparable with UTM and SMPO. For a window length of 20 ms (640 samples), the proposed method requires 2.66 s where STMT takes 27466.80 s, UTM takes 0.46 s and SMPO takes 0.22 s. This ultrafast performance of the proposed method, where the process time is significantly less than the speech length, allows us to perform the speech enhancement in a real time environment.

Table 5: Mean Scores of BAK scale for different methods in presence of Gaussian white noise at 5 db

| Listener | UTM | SMPO | STMT | Proposed Method |
|---|---|---|---|---|
| 1 | 4.0 | 4.5 | 5.0 | 5.0 |
| 2 | 4.3 | 4.9 | 4.9 | 4.7 |
| 3 | 4.2 | 4.4 | 4.8 | 4.9 |
| 4 | 4.4 | 4.7 | 4.9 | 4.8 |
| 5 | 4.2 | 4.8 | 4.8 | 4.7 |
| 6 | 3.9 | 4.6 | 4.7 | 4.9 |
| 7 | 3.8 | 3.9 | 4.3 | 4.4 |
| 8 | 4.4 | 4.6 | 4.7 | 4.6 |
| 9 | 3.5 | 3.8 | 4.3 | 4.5 |
| 10 | 4.2 | 4.5 | 4.7 | 4.8 |

Table 6: Mean Scores of OVL scale for different methods in presence of Gaussian white noise at 5 db

| Listener | UTM | SMPO | STMT | Proposed Method |
|---|---|---|---|---|
| 1 | 2.6 | 4.0 | 4.2 | 4.1 |
| 2 | 3.3 | 3.8 | 4.1 | 3.7 |
| 3 | 3.9 | 4.1 | 4.4 | 4.3 |
| 4 | 3.6 | 4.2 | 4.3 | 4.2 |
| 5 | 3.3 | 3.9 | 4.0 | 4.1 |
| 6 | 3.9 | 4.6 | 4.8 | 4.9 |
| 7 | 3.8 | 3.8 | 4.5 | 4.3 |
| 8 | 3.6 | 4.1 | 4.2 | 4.2 |
| 9 | 3.5 | 4.5 | 4.6 | 4.7 |
| 10 | 3.9 | 4.6 | 4.9 | 4.8 |

Table 7: Mean Scores of SIG scale for different methods in presence of street noise at 5 db

| Listener | UTM | SMPO | STMT | Proposed Method |
|---|---|---|---|---|
| 1 | 3.6 | 4.0 | 4.1 | 4.0 |
| 2 | 3.3 | 3.9 | 3.8 | 3.7 |
| 3 | 3.9 | 4.0 | 4.3 | 4.2 |
| 4 | 3.4 | 4.2 | 4.2 | 4.5 |
| 5 | 3.2 | 3.8 | 4.1 | 4.0 |
| 6 | 2.9 | 3.6 | 3.8 | 3.9 |
| 7 | 3.8 | 3.8 | 4.3 | 4.2 |
| 8 | 3.4 | 3.6 | 4.3 | 4.1 |
| 9 | 3.5 | 3.9 | 3.8 | 3.7 |
| 10 | 3.7 | 3.8 | 3.8 | 3.9 |

Table 8: Mean Scores of BAK scale for different methods in presence of street noise at 5 db

| Listener | UTM | SMPO | STMT | Proposed Method |
|---|---|---|---|---|
| 1 | 4.0 | 4.5 | 5.0 | 5.0 |
| 2 | 4.3 | 4.9 | 4.9 | 4.7 |
| 3 | 4.2 | 4.4 | 4.7 | 4.9 |
| 4 | 4.4 | 4.7 | 4.6 | 4.8 |
| 5 | 4.2 | 4.8 | 4.8 | 4.7 |
| 6 | 3.9 | 4.6 | 4.8 | 4.9 |
| 7 | 3.8 | 3.9 | 4.5 | 4.4 |
| 8 | 4.4 | 4.6 | 4.6 | 4.7 |
| 9 | 3.5 | 3.9 | 4.2 | 4.7 |
| 10 | 4.7 | 4.8 | 4.8 | 4.9 |

Table 9: Mean Scores of OVL scale for different methods in presence of street noise at 5 db

| Listener | UTM | SMPO | STMT | Proposed Method |
|---|---|---|---|---|
| 1 | 2.6 | 4.0 | 4.1 | 4.1 |
| 2 | 3.3 | 3.8 | 3.9 | 3.7 |
| 3 | 3.9 | 4.1 | 4.2 | 4.3 |
| 4 | 3.6 | 4.2 | 4.1 | 4.2 |
| 5 | 3.3 | 3.9 | 4.2 | 4.1 |
| 6 | 3.9 | 4.6 | 4.8 | 4.9 |
| 7 | 3.8 | 3.8 | 4.2 | 4.3 |
| 8 | 3.6 | 4.1 | 4.3 | 4.2 |
| 9 | 3.5 | 4.5 | 4.4 | 4.7 |
| 10 | 3.9 | 4.8 | 4.9 | 4.9 |

## 4. Conclusions

In this paper, for a real time speech processing application, in order to obtain a suitable threshold value for thresholding operation in a subband of PWP transformed noisy speech, we have developed a threshold determination technique based on modeling of the TE operated PWP coefficients of the noisy speech and noise by Erlang-2 PDF. Unlike the other wavelet packet based thresholding methods, the proposed method is very fast and allows real time speech processing. The threshold value obtained in the proposed method is adaptive to the speech and silence subbands. The PWP coefficients of the noisy speech are thresholded by exploiting the derived threshold in the proposed custom thresholding function, which works as a $\mu$-law or a semisoft thresholding function or their combination based on the probability of speech presence and absence in a subband. Extensive simulations and results attest that the proposed method produces better enhanced speech with higher objective metrics, namely Segmental SNR Improvement, PESQ, and WSS values in comparison to those of the competing methods. Much better spectrogram outputs and the higher scores in the formal subjective listening tests are also presented as the indicators of improved performance of the proposed method.

Table 10: Computational time for different algorithms

| Window size (ms) | Method | Time (s) |
|---|---|---|
| 20 | UTM | 0.46 |
| | SMPO | 0.22 |
| | STMT | 27466.80 |
| | Proposed method | 2.66 |
| 40 | UTM | 0.46 |
| | SMPO | 0.17 |
| | STMT | 14009.34 |
| | Proposed method | 1.52 |
| 60 | UTM | 0.46 |
| | SMPO | 0.14 |
| | STMT | 9465.22 |
| | Proposed method | 1.21 |
| 80 | UTM | 0.46 |
| | SMPO | 0.12 |
| | STMT | 6124.51 |
| | Proposed method | 0.81 |

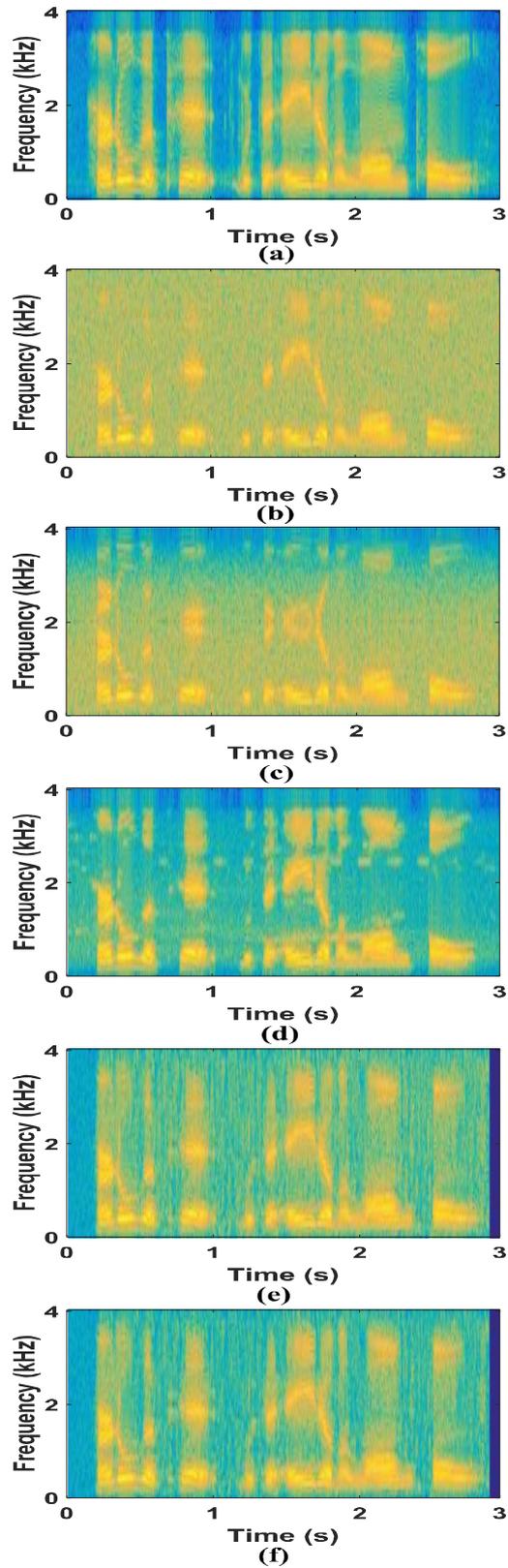

Figure 18: Spectrograms of (a) clean signal (b) noisy signal with 10dB Gaussian white noise; spectrograms of enhanced speech from (c) UTM (d) SMPO (e) STMT (f) proposed method

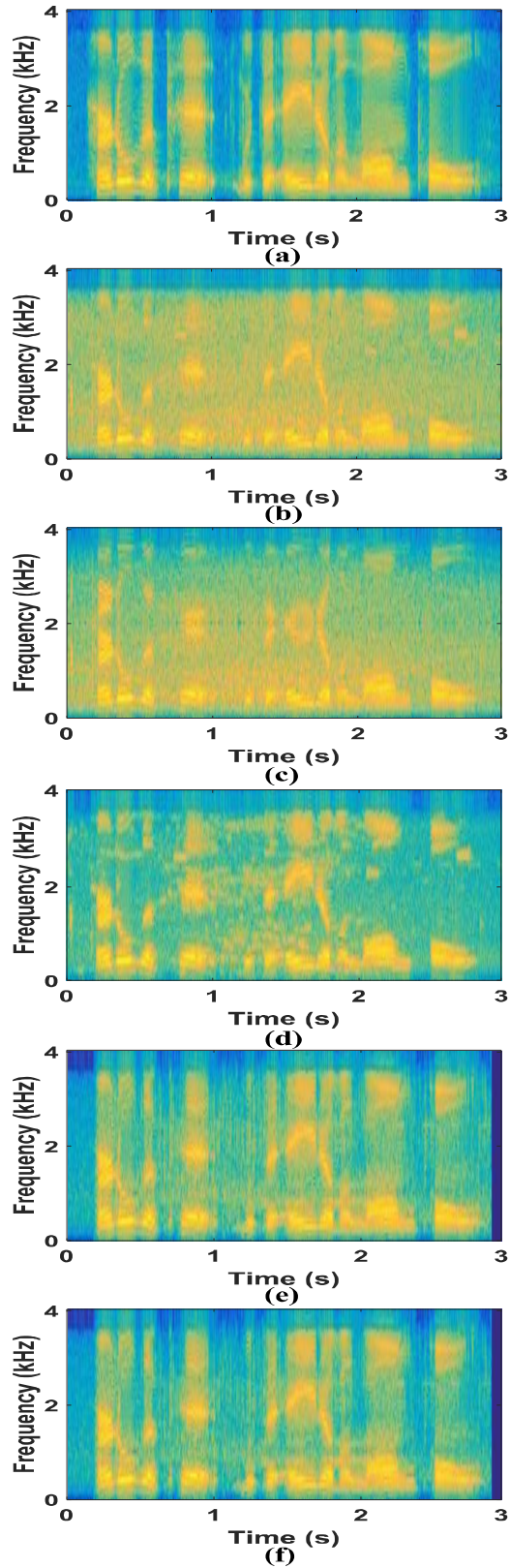

Figure 19: Spectrograms of (a) clean signal (b) noisy signal with 10dB street noise; spectrograms of enhanced speech from (c) UTM (d) SMPO (e) STMT (f) proposed method